# Dynamics of COVID-19 Misinformation: An Analysis of Conspiracy Theories, Fake Remedies, and False Reports

Nirmalya Thakur[1], Mingchen Shao[2], Victoria Knieling[3], Vanessa Su[2], Andrew Bian[4], and Hongseok Jeong[2]

[1] Department of Electrical Engineering and Computer Science, South Dakota School of Mines and Technology, Rapid City, SD 57701, USA
[2] Department of Computer Science, Emory University, Atlanta, GA 30322, USA
[3] Program in Linguistics, Emory University, Atlanta, GA 30322, USA
[4] Goizueta Business School, Emory University, Atlanta, GA 30322

`nirmalya.thakur@sdsmt.edu, katie.shao@emory.edu, victoria.knieling@emory.edu,vanessa.su@emory.edu,andrew.bian@emory.edu,peter.jeong@emory.edu`

**Abstract.** This paper makes four scientific contributions to the area of misinformation detection and analysis on digital platforms, with a specific focus on investigating how conspiracy theories, fake remedies, and false reports emerge, propagate, and shape public perceptions in the context of COVID-19. A dataset of 5,614 posts on the internet that contained misinformation about COVID-19 was used for this study. These posts were published in 2020 on 427 online sources (such as social media platforms, news channels, and online blogs) from 193 countries and in 49 languages. First, this paper presents a structured, three-tier analytical framework that investigates how multiple motives - including fear, politics, and profit - can lead to a misleading claim. Second, it emphasizes the importance of narrative structures, systematically identifying and quantifying the thematic elements that drive conspiracy theories, fake remedies, and false reports. Third, it presents a comprehensive analysis of different sources of misinformation, highlighting the varied roles played by individuals, state-based organizations, media outlets, and other sources. Finally, it discusses multiple potential implications of these findings for public policy and health communication, illustrating how insights gained from motive, narrative, and source analyses can guide more targeted interventions in the context of misinformation detection on digital platforms.

## 1 Introduction

The first cases of the COVID-19 pandemic, caused by the SARS-CoV-2 virus, were recorded in a seafood market in Wuhan, China, in December 2019 [1,2]. Over the next few months, COVID-19 spread to different parts of the world at an unprecedented rate, prompting the World Health Organization (WHO) to classify COVID-19 as a pandemic on March 11, 2020 [3]. With no established treatments or vaccines in the early stages [4], many countries resorted to strict lockdowns and related measures to slow the spread of the virus [5-8]. These policies spurred disruptive shifts in virtually every segment of society, especially in education, where schools and universities scrambled to transition instructional delivery, assessments, and support services onto virtual platforms [9]. This forced adaptation highlighted the fragility of global systems under crisis conditions, as entire populations started to depend on digital tools not only for academic continuity but also for basic information, social engagement, and a sense of community [10]. Amid these challenges, digital platforms emerged as primary sources for public discourse, delivering real-time updates on infection rates, travel advisories, emerging scientific evidence, and guidelines recommended by governments and policy-making bodies [11,12]. Although these sources provided vital information to the general public, they also served as mediums for rumors, fake news, and speculative claims [13], leading to the creation and dissemination of misinformation [14,15].

The global proliferation of misinformation since the beginning of the COVID-19 pandemic has generated profound social and political challenges, prompting urgent calls for innovative research and solutions. As communities worldwide grapple with navigating legitimate information related to health, vaccine safety, and government policies, misinformation continues to circulate at an alarming rate [16-18]. Recent studies emphasize the rapidity with which fake news spreads through both traditional media and social media platforms, often outpacing fact-based reporting [19,20]. Such a phenomenon not only undermines public trust in governments and policy-making bodies but can also trigger dangerous responses - ranging from the rejection of credible health guidelines to the rise of conspiracy-driven protests [21]. Despite the growing body of literature on misinformation detection and analysis in the context of COVID-19, many existing works primarily focus on lexical or sentiment-based classification, overlooking deeper structural factors such as motives, content narratives, and sources. This research gap is significant, given the rapid rate of generation of fake news and the complexities involved in predicting and tracking misinformation on digital platforms [22-24].

In response to the pandemic's "infodemic", researchers in this field have investigated several approaches to analyze and combat misinformation. For instance, some have utilized machine-learning techniques to classify textual content based on sentiment polarity [25]. Others have applied network analysis to uncover the role of social bots and influencer accounts in accelerating the spread of sensationalized or politically driven narratives [26]. While these studies have yielded valuable insights into the broader patterns of misinformation on digital platforms, most of these works neglect the examination of how underlying motives - such as fear, profit, or political agendas - intersect with particular themes and narrative structures. The critical interplay between what spurs misinformation actors to take action and how they shape the messaging around conspiracy theories, false reports, or fake remedies remains underexplored [27]. Recent papers highlight this need by noting that a purely topical approach to misinformation fails to highlight the strategic intent behind its creation and dissemination [28-30]. Addressing this gap is crucial to designing targeted interventions that can tackle the root causes of misinformation on digital platforms.

A few recent works in this field have analyzed the narrative structures that drive misinformation campaigns [31]. Narratives that misrepresent scientific data, distort historical facts, or invoke conspiracy theories can have considerable influence over specific populations, often rooted in sociopolitical or cultural factors [32]. However, the literature in this area of research is still evolving in terms of systematically categorizing these narratives across different misinformation types. While a few works have offered preliminary taxonomies of COVID-19 misinformation narratives - covering themes like unverified cures, virus origins, and governmental responses - they frequently treat these narratives as broad categories without nuanced analysis of their relative prevalence or shifts over time [33-36]. More granular insights into how narratives adapt to different platforms, linguistic contexts, and audience needs are still largely missing. Existing studies also rarely integrate the narrative approach with comprehensive motive analysis, thereby missing an opportunity to enrich understanding of both the "why" and the "how" behind misinformation campaigns.

Simultaneously, there have been multiple studies that recognize the critical role of misinformation sources, yet many investigations focus almost exclusively on the influence of social media platforms. While some researchers have highlighted how individual users, organized groups, and mainstream or fringe media contribute to the creation of misinformation [37-39], none of these works have investigated the role different sources play in the context of conspiracy theories, fake remedies, and false reports. Quantifying and comparing the role and influence of different types of sources in the context of conspiracy theories, fake remedies, and false reports is vital for developing fact-checking strategies and platform-level interventions. These interventions should not only address the presence of misinformation but also consider how the networks of individuals, media outlets, and other organizations interact and contribute to its spread [40,41]. The literature in this field lacks a comprehensive framework that highlights the roles of motives, narratives, and sources to reveal a more holistic picture of the misinformation ecosystem. Addressing these research gaps serves as the main motivation for this study.

The work of this paper focuses on investigating and analyzing three critical areas in the context of COVID-19-related misinformation research: (1) the motives that drive the creation and dissemination of misinformation, (2) the main narratives used across different types of misinformation, such as conspiracy theories, false reports, and fake remedies, and (3) the key sources or actors responsible for spreading these narratives. We build on prior studies that have used data-driven methods to examine specific misinformation attributes while aiming to augment their findings through a more comprehensive approach [42,43]. Our study stands out in its emphasis on categorizing multiple motives and narratives across different types of misinformation, such as conspiracy theories, false reports, and fake remedies, thereby revealing complex overlaps - such as how profit-driven and fear-based agendas may intersect in this context. We also present a comprehensive analysis of how different sources - individuals, media organizations, politicians, and others - collaborate or operate independently to propagate misinformation [44-47]. By quantifying each source's contribution and linking these contributions to specific content types, our approach highlights patterns that were previously not investigated in the broader, predominantly content-focused studies in this field. In combining these dimensions of misinformation analysis, our research offers a novel perspective that is expected to support both digital policy interventions [48,49] and automated systems designed to detect and contain misinformation [50,51]. By systematically examining why misinformation is spread, what themes are mainly represented, and what sources play crucial roles in its dissemination, this study addresses critical research gaps in the field of misinformation research. The analytical framework we propose, supported by empirical findings, offers a template other researchers can adapt to investigate similar research problems arising in various public health and political contexts.

The remainder of this paper is organized as follows. First, the methodology section begins with an overview of the dataset used in this study. Thereafter, the rest of that section discusses how the dataset was pre-processed to interpret and analyze different types of motives, narratives, and sources. More specifically, we explain the development of our multi-step analytical framework, highlighting the computational techniques involved in generating quantitative and qualitative insights. In this section, we also discuss the time complexity and space complexity of our approach. The next section presents the results related to the three critical areas of COVID-19 misinformation research that we focused on in this study. In the results section, we also discuss the implications of these findings, highlighting how they can guide future strategies for mitigating the influence of misinformation. Finally, the conclusions section summarizes the contributions of our paper and outlines the future scope of work.

## 2 Methodology

The dataset used for this work was developed by Shapiro et al. [52]. This dataset presents a comprehensive collection of multiple cases of COVID-19 misinformation, each entry capturing important details related to the nature, source, and intent of the content. To develop this dataset, Shapiro et al. [52] collected data from various publicly accessible outlets, including social media, online news portals, and blogs. This dataset consists of 5,614 online posts containing misinformation about COVID-19. These posts were published in 2020 across 427 online sources, including social media platforms, news channels, and blogs, covering 193 countries and 49 languages. Each record in this dataset represents a single instance of COVID-19-related misinformation that includes the specific URL or other point of reference through the "Reported_On" column, along with the "Title" of the post. The publication data and language of the post are presented in the "Publication_Date" and "Primary_Language" columns, respectively. In addition to these contextual fields, the dataset applies a classification scheme that groups misinformation into three broad categories – conspiracy theories, fake remedies, and false reports. The results of this classification are presented in the "Misinfo_Type" attribute in the dataset. The "Main_Narrative" attribute elaborates on the main theme of each

record - such as accounts of virus weaponization or unverified cures - and is supplemented by a "Narrative_Description" attribute, which elaborates on the nuances and potential ramifications of the misinformation. Every record also contains a "Motive" attribute, reflecting why a particular claim might have been disseminated (whether due to fear, political gain, profit, hope, etc.), and a "Source" descriptor, indicating whether the content originated from individuals, media outlets, politicians, corporations, or other entities. At the time of writing this paper, as per the best knowledge of the authors, no other dataset in this field presents COVID-19-related misinformation along with these characteristics. So, this dataset was selected for this research project.

The program that governs our analytical framework was developed to systematically evaluate the dataset's content and interpret how different types and instances of misinformation are created and disseminated. All these programs were written in Python 3.10 - installed on a computer with a Microsoft Windows 10 Pro operating system. Prior to performing any analysis, we performed comprehensive data pre-processing of this dataset. This step included ensuring that the crucial text-based columns ("Motive", "Misinfo_Type", "Main_Narrative", and "Source") were all standardized. Rows with missing values were also dropped during this data pre-processing step.

After completing the data pre-processing, we wrote a program in Python that used a three-step approach for investigating and analyzing the three critical areas in the context of COVID-19-related misinformation that this study focuses on. First, it analyzed the relationship between "Motive" and "Misinfo_Type", generating raw counts to reveal which motives occur most frequently within each misinformation type. Conspiracy theories, for example, might exhibit high levels of politically charged or fear-driven motives [53], while false remedies could rely more on confidence in unproven cures [54]. False reports, on the other hand, might capture an interplay of fear, political messaging, and sporadic commercial incentives [55]. So, the program also calculated percentages to deduce a clearer interpretation of each motive's relative weight.

In the second step, the program focused on analyzing the "Main_Narrative" column. Conspiracy theories might comprise catchy storylines claiming incorrect origins of the virus [56], false reports might present incorrect updates on COVID-19 cases written appealingly [57], and fake remedies might promote untested cures and therapies related to COVID-19 [58]. So, in this step, we identified the prominent and leading themes that defined conspiracy theories, fake remedies, and false reports. This step aimed to infer which storylines yielded the strongest influence across different misinformation categories.

In the third step, the program identified the most frequent sources for each type of misinformation. By analyzing the "Source" column, this step aimed to present insights into the roles played by individuals, corporate entities, politicians, media outlets, or other bodies in propagating each form of COVID-19-related misinformation. Algorithm 1 presents the pseudocode of this program, which is followed by an analysis of the time complexity and space complexity of this approach.

| Algorithm 1: Misinformation Analysis and Interpretation |
|---|
| **Input**: Pre-processed Dataset |
| **Output**: Three figures illustrating: |
| Aggregated motive distributions across all misinformation types |
| Aggregated narrative distributions across all misinformation types |
| Aggregated top sources across all misinformation types |
| |
| **Local Variables**: dataFrame, motiveCounts, narrativeCounts, topSources, motiveTotals, narrativeTotals, imagesDir |
| |
| **def AnalyzeMotives**(dataFrame) |
|    motiveCounts ← default dictionary of (misinformationType → dictionary of (motive → count)) |
|    for each row in dataFrame do: |
|      if 'Misinfo_Type' not null AND 'Motive' not null then: |
|        typesList ← split row['Misinfo_Type'] by comma |
|        for each misType in typesList do: |
|          motiveCounts[misType][row['Motive']] ← motiveCounts[misType][row['Motive']] + 1 |
|        end of for |
|      end if |
|    end of for |
|    return motiveCounts |
| end of function |
| |
| **def AnalyzeNarratives**(dataFrame) |
|    narrativeCounts ← default dictionary of (misinformationType → dictionary of (narrative → count)) |
| |
|    for each row in dataFrame do: |
|      if 'Misinfo_Type' not null AND 'Main_Narrative' not null then: |
|        typesList ← split row['Misinfo_Type'] by comma |
|        for each misType in typesList do: |
|          narrativeCounts[misType][row['Main_Narrative']] ← narrativeCounts[misType][row['Main_Narrative']] + 1 |
|        end of for |
|      end if |
|    end of for |
|    return narrativeCounts |
| end of function |
| |
| **def GetTopSources**(filteredData) |
|    sourceCountSeries ← filteredData['Source'].value_counts() |
|    sourceCountSeries ← pick the largest 10 from sourceCountSeries |
|    totalCount ← sum of all counts in sourceCountSeries |
|    topSourcesDict ← empty dictionary |
|    for each (source, countValue) in sourceCountSeries do: |
|      percentageVal ← (countValue / totalCount) × 100 |
|      topSourcesDict[source] ← (countValue, percentageVal) |
|    end of for |
|    return topSourcesDict |
| end of function |

```
def AnalyzeSources(dataFrame)
    topSources ← empty dictionary
    misinfoTypesList ← ['conspiracy', 'fake remedy', 'false reporting']
    for each misType in misinfoTypesList do:
        filteredDF ← rows in dataFrame where 'Misinfo_Type' contains misType (case-insensitive)
        topSources[misType] ← GetTopSources(filteredDF)
    end of for
    return topSources
end of function

def CalculateNormalizationTotals(motiveCounts, narrativeCounts)
    motiveTotals ← empty dictionary
    narrativeTotals ← empty dictionary
    for each misType in motiveCounts do:
        totalMotiveCount ← sum of all counts in motiveCounts[misType]
        motiveTotals[misType] ← totalMotiveCount
    end of for
    for each misType in narrativeCounts do:
        totalNarrativeCount ← sum of all counts in narrativeCounts[misType]
        narrativeTotals[misType] ← totalNarrativeCount
    end of for
    return motiveTotals, narrativeTotals
end of function

def GenerateFigures(motiveCounts, narrativeCounts, topSources, imagesDir)
    if folder imagesDir does not exist:
        make directory imagesDir
    end if
    plot figure with top motives per misinformation type
    plot figure with top narratives per misinformation type
    plot figure with top 10 sources per misinformation type
    save each figure as a .jpg in imagesDir
end of function
```

This program begins by reading data from the pre-processed dataset. As each row in the file must be processed, this operation takes O(N) time, where N is the number of rows. At the time of analyzing the motives associated with each misinformation type, the program once again iterates through every row, that takes O(N) time. Within each row, it may split comma-separated strings, leading to an additional factor of O(K), where K is the average number of misinformation types per row. As each individual update to a 'defaultdict' is O(1), the overall complexity of motive analysis is O(NK). A similar process applies to analyzing narratives, so that step likewise has a complexity of O(NK). To identify the top ten sources, the program uses 'value_counts()', which operates in O(N). Since 'nlargest(10)' selects a small subset, that portion is effectively O(1). Filtering for each misinformation type is O(N), because it must check all rows. Consequently, each misinformation-type-specific filtering phase is O(N), repeated for however many types exist. For presenting the results visually, the sorting required for plots is O(M log M), where M represents the number of unique elements being sorted

(for instance, unique motives or narratives). Therefore, the time taken for the visualization can be represented as O(M log M). As K and T (the number of misinformation types) are usually small constants, the most substantial terms in overall time complexity are O(NK) and O(M log M). As a result, we can approximate the total complexity as O(N + M log M), indicating that data size (N) and the scale of unique categories (M) both play important roles in runtime performance.

From a storage standpoint, the dataset itself takes up O(NM) memory, where N is the number of rows and M is the number of columns. In addition, dictionaries used for counting motives and narratives (e.g., 'motive_counts' and 'narrative_counts') each occupy O(MT), where T is the number of misinformation types, and M is the number of unique items in that category. The top ten sources for each type are stored in O(TS), with S denoting how many sources are retained per type. During plotting, 'matplotlib' must keep figure data in memory, incurring about O(M) space usage. Combining these requirements yields a total space usage of O(NM) + O(MT) + O(M). As T remains small, the dominant cost typically remains O(NM), governed mainly by the size of the original dataset.

## 3    Results and Discussions

This section presents the results and findings of this research project. The findings from our analysis reveal the complex nature of COVID-19 misinformation creation and dissemination - one in which motives, narratives, and sources dynamically interact to spread conspiracy theories, fake remedies, and false reports. By examining each characteristic in detail, our work goes beyond basic categorizations and focuses on how specific motivations and dissemination pathways interact with one another. In the rest of this section, we present a comprehensive discussion of these results, highlighting the novel findings as well as the broader implications of these results.

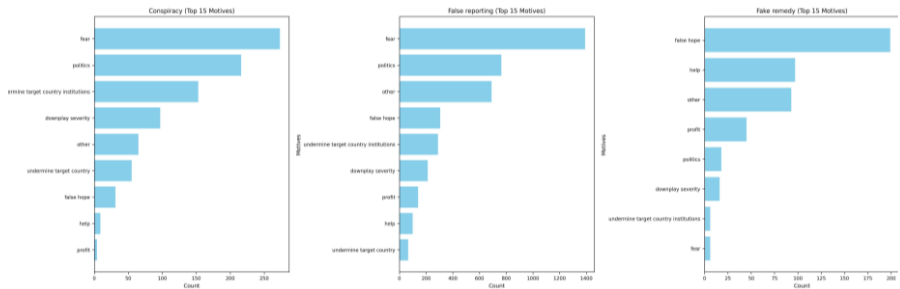

**Fig. 1.** An analysis of motives by misinformation type.

A key finding of this study reveals the distinct motives that lead to the creation of different types of COVID-19 misinformation, as shown in Figure 1. Within conspiracy theories, fear accounts for roughly 30.23% of the overall motive distribution, surpassing all other motives. Although it is unsurprising that fear features prominently in conspiracy discourse, particularly amid an unfolding global crisis - our findings confirm the potency of fear as a catalyst for the rapid uptake of such theories [59,60]. Coupled with political motivations (23.92%) and systematic efforts to undermine trust in institutions (16.94%), these fear-based narratives may exploit the anxieties of the general public to spread distrust in governments or global organizations. Notably, the interplay between fear and politics differentiates conspiracy theories from other misinformation types. The situation changes somewhat when we turn to false reporting, where the proportion of fear-based content rises even further to 35.16%. This increment, though modest, indicates that fear is highly effective at drawing public attention to false stories about the pandemic. Beyond fear, our findings showed a broader distribution of motives - political agendas (19.30%), efforts to undermine institutional credibility (7.31%), the spread of false hope (7.72%), and even occasional attempts to help (2.50%). This broader spectrum suggests that those fabricating or disseminating false reports tailor their messaging to various audience segments, offering content that resonates with people's pre-existing concerns and curiosities. Furthermore, our findings highlight that fear and politics often work hand in hand, although in slightly different proportions than those seen in conspiracy theories. By quantifying how these motives shift across misinformation types, our approach offers a granular view of how individuals responsible for false reporting tweak their messaging to maximize reach or compliance. A considerably different pattern was observed for fake remedies. Here, the high-level motive was false hope, dominating at approximately 41.46%. The pandemic's urgency has understandably spurred a widespread desire for quick fixes [61], and our findings suggest that many individuals - perhaps out of desperation - are inclined to believe in and share unverified treatments. Our findings also showed that a significant minority of fake remedy posts stem from attempts to help (20.21%), suggesting that some individuals disseminate such claims because they genuinely believe in unconventional therapies, even if those therapies lack scientific evidence. These findings support the findings of prior works that misinformation cannot always be attributed to ill intent [62-66]. Profit-driven motives make up about 9.38%, showing that some businesses or individuals may be taking advantage of public fears to sell unreliable treatments.

Equally interesting are our findings on misinformation narratives and the themes used to capture audience interest, as shown in Figure 2. For conspiracy theories, nearly 20% of the narratives suggest that COVID-19 is a weaponized or artificially designed agent, and they often coexist with related themes challenging the nature (13.95%) or origin (11.52%) of the virus. Government responses account for about 10.52% of conspiracy narratives, reflecting a persistent skepticism toward official actions - such as lockdown policies [67], border closures [68], or vaccine distribution plans [69]. By distinguishing each of these narratives and quantifying their respective frequencies, our findings show how conspiracy theorists shaped their messages to be simultaneously sensational, fear-inducing, and politically charged. False reporting, while often regarded as more mundane than conspiracies [70], indicated multiple attention-grabbing themes. Emergency responses ranked at 21.80%, mirroring the public's consistent focus on urgent governmental actions, hospital capacities, and rescue operations. Similarly, stories regarding the COVID-19 status of individuals made up 17.80% of the themes, presumably fueled by personal accounts of high-profile or everyday people contracting the virus. Though government responses (11.81%) also feature prominently, this category broadens to include a wide range of subjects. Fake remedies, by contrast, were dominated by a single overarching theme: false cures and preventative measures. At 93.12%, the near uniformity of this narrative indicated a clear pattern of exploiting individuals' urgent desire to shield themselves and their loved ones from the virus. Although marginal topics such as emergency responses (1.25%) and false diagnostic procedures (1.67%) appeared, they were much less compared to the scale of claims about rapid, miraculous cures. This highlights how strongly fake remedies resonated with a global need for a cure or prevention against COVID-19. By highlighting how this emphasis on fake remedies dwarfs all other narratives in this category, our findings show that fake remedies have a unique thematic structure among other COVID-19 misinformation types.

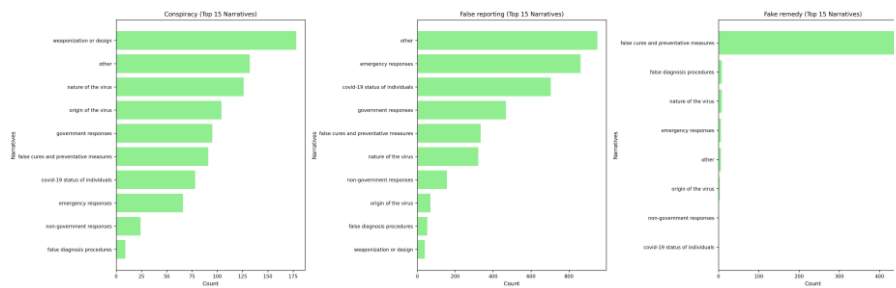

**Fig. 2.** An analysis of narratives by misinformation type.

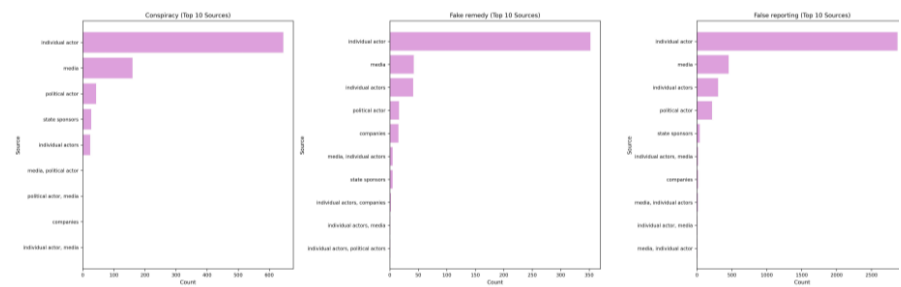

**Fig. 3.** An analysis of sources by misinformation type.

In parallel to these narrative findings, our study also highlights the channels of dissemination, pinpointing those most responsible for creating and disseminating different types of misinformation, as shown in Figure 3. For conspiracy theories, individuals were the largest contributor, at approximately 71.43%. This pattern suggests that decentralized networks - often operating on social media or private messaging platforms, can significantly spread conspiracy theories, outpacing even established media outlets (17.72%) or politicians (4.76%). For fake remedies, individuals (73.33%) again featured as the top misinformation source, but they coexisted with a more diverse cast of media (8.75%), group-level individuals (8.54%), and even for-profit companies

(3.12%). The presence of companies indicates that financially motivated interests may seek to promote untested cures to generate revenue [71]. Meanwhile, the involvement of media outlets, though not as dominant as individuals, underscores how some fringe or sensationalist media channels might exploit health crises [72]. False reporting emerged with a similar pattern: individuals again led at 72.86%, with media (11.52%) and group-level individuals (7.71%) in second and third place. While politicians (5.48%) and state-sponsored sources (1.04%) also contributed, these numbers, though relatively modest, still carry weight in influencing larger narratives around pandemic policy and management. The notion that decentralized online communities can rival or surpass official channels in terms of dissemination speed is highlighted here, pointing to the tremendous ease with which fabricated stories travel once they trigger individual anxieties.

These insights are expected to have real-world applications in multiple contexts. Public health organizations, for instance, may rely on this granular breakdown of motives and narratives to craft nuanced advisories that address the root causes behind different types of misinformation. If fear emerges as the primary driver in certain scenarios, more empathic and transparent communication strategies could help lessen the inclination to accept or share misinformation. Meanwhile, policymakers concerned about political manipulation could set up targeted oversight measures [73] for sources identified as especially active in pushing politicized conspiracy claims. Similarly, commercial platforms can adapt their content moderation strategies [74] to recognize and flag emerging conspiracy themes or false cure narratives before they scale. By incorporating knowledge of overlapping motives and cross-category narratives, social media sites may detect new types of misinformation that do not match prior patterns but still follow recognizable rhetorical or motivational cues.

These findings may also be applied in journalism and fact-checking networks. Based on a detailed taxonomy of how fear, politics, and profit come together in misinformation posts, fact-checkers may categorize false claims more swiftly and focus on the stories that have the most impact [75]. Real-time collaboration with social media platforms could then slow the viral spread of high-risk misinformation clusters, especially those presenting immediate public health dangers. Over time, the identification of repeating narrative frames - such as repeated claims of miracle cures, could serve as a valuable early-warning system [76-79], allowing fact-checkers and governmental agencies to intervene sooner. Education and training initiatives, both for journalists and the public, could incorporate scenario-based learning [80] that mirrors the patterns revealed here, teaching stakeholders to scrutinize content more critically.

In addition, the findings on how multiple stakeholders - ranging from individuals and small collectives to official media outlets - intermingle in spreading misinformation are expected to inform broader social and policy-oriented interventions. Collaboration between technology platforms and civil society organizations may prove fruitful, as they can allocate their combined resources to address specific behaviors. If, for example, individuals are sharing conspiracy theories in private chat groups, an educational campaign or improved platform-level policy could reduce the perceived credibility of those narratives. At a legislative level, policymakers may use these insights to formulate more balanced guidelines that address the potential for political and corporate misuse of crisis narratives, all while maintaining open discourse [81].

Although our results present multiple novel findings related to the creation and dissemination of COVID-19-related misinformation, this paper has multiple limitations. First, the findings presented in this paper are based on the dataset that was used. This dataset, while comprehensive, does not present every source of COVID-19-related misinformation, particularly in regions or languages not represented in the data. Second, we did not verify the labels assigned to the different sources of misinformation in this dataset. Finally, misinformation evolves rapidly, frequently responding to shifts in current events, public sentiment, and politics. Given these dynamics, the trends reported in this paper may change as new variants of the virus or new public policies emerge, prompting fresh narratives or motives.

## 4    Conclusions

The findings of this paper underscore the importance of approaching COVID-19 misinformation through different interrelated perspectives. Rather than treating misinformation as a single-layer phenomenon - such as limiting the focus to topic classification alone - our investigation integrates motives, narrative elements, and dissemination channels to highlight their role in this context. Our findings indicate that fear, financial gain, and political agendas sometimes operate in parallel, giving conspiracy theories or fake remedies a potent appeal. Equally significant is that individuals, media organizations, and corporate entities often amplify these narratives, which may be intentional or unintentional. By illustrating how motives, narratives, and sources can intertwine, this work provides a stronger empirical foundation for those seeking to recognize, analyze, and mitigate the negative effects of misinformation.

Future research could build on this framework in multiple ways. A direct extension would be to track misinformation trends over time, examining how dominant motives and narratives shift in response to changing social, political, or epidemiological contexts. Such a longitudinal perspective might clarify whether certain false claims fade as public knowledge evolves or merely transform themselves to remain relevant. Additionally, there is scope to expand these methods to different languages and cultural contexts, providing evidence on whether core themes, emotional appeals, or sources differ substantially across regions. Another extension could be integrating advanced tools, including network analysis or machine learning, to identify broader patterns of coordination among different sources of misinformation. By pursuing these lines of inquiry, researchers, platform developers, and policy experts may gain an even deeper understanding of misinformation dynamics and refine their strategies to contain its spread in an ever-changing digital world.

**Disclosure of Interests.** The authors have no competing interests to declare that are relevant to the content of this article.

# References

1. Ciotti, M., Ciccozzi, M., Terrinoni, A., Jiang, W.-C., Wang, C.-B., Bernardini, S.: The COVID-19 pandemic. Crit. Rev. Clin. Lab. Sci. 57, 365–388 (2020). https://doi.org/10.1080/10408363.2020.1783198.
2. Velavan, T.P., Meyer, C.G.: The COVID-19 epidemic. Trop. Med. Int. Health. 25, 278–280 (2020). https://doi.org/10.1111/tmi.13383.
3. Cucinotta, D., Vanelli, M.: WHO declares COVID-19 a pandemic. Acta Biomed. 91, 157–160 (2020). https://doi.org/10.23750/abm.v91i1.9397.
4. Stasi, C., Fallani, S., Voller, F., Silvestri, C.: Treatment for COVID-19: An overview. Eur. J. Pharmacol. 889, 173644 (2020). https://doi.org/10.1016/j.ejphar.2020.173644.
5. Allen, D.W.: Covid-19 lockdown cost/benefits: A critical assessment of the literature. Int. J. Econ. Bus. 29, 1–32 (2022). https://doi.org/10.1080/13571516.2021.1976051.
6. Herby, J., Jonung, L., Hanke, S.H.: A systematic literature review and meta-analysis of the effects of lockdowns on COVID-19 mortality II, http://dx.doi.org/10.1101/2023.08.30.23294845, (2023). https://doi.org/10.1101/2023.08.30.23294845.
7. Le, K., Nguyen, M.: The psychological consequences of COVID-19 lockdowns. In: The Political Economy of Covid-19. pp. 39–55. Routledge, London (2022).
8. Greyling, T., Rossouw, S., Adhikari, T.: The good, the bad and the ugly of lockdowns during Covid-19. PLoS One. 16, e0245546 (2021). https://doi.org/10.1371/journal.pone.0245546.
9. Thakur, N.: A large-scale dataset of Twitter chatter about online learning during the current COVID-19 Omicron wave. Data (Basel). 7, 109 (2022). https://doi.org/10.3390/data7080109.
10. Thakur, N., Han, C.: An exploratory study of tweets about the SARS-CoV-2 Omicron variant: Insights from sentiment analysis, language interpretation, source tracking, type classification, and embedded URL detection. COVID. 2, 1026–1049 (2022). https://doi.org/10.3390/covid2080076.
11. Bernardino, M., Bacelar Nicolau, L.: The importance of reliable social media information during the COVID-19 pandemic. Eur. J. Public Health. 30, (2020). https://doi.org/10.1093/eurpub/ckaa165.067.
12. Thakur, N., Khanna, K., Cui, S., Azizi, N., Liu, Z.: Mining and analysis of search interests related to online learning platforms from different countries since the beginning of COVID-19. In: Lecture Notes in Computer Science. pp. 280–307. Springer Nature Switzerland, Cham (2023).
13. Ansani, A., Marini, M., Cecconi, C., Dragoni, D., Rinallo, E., Poggi, I., Mallia, L.: Analyzing the perceived utility of Covid-19 countermeasures: The role of pronominalization, Moral Foundations, moral Disengagement, Fake News embracing, and Health Anxiety. Psychol. Rep. 125, 2591–2622 (2022). https://doi.org/10.1177/00332941211027829.
14. Solomon, D.H., Bucala, R., Kaplan, M.J., Nigrovic, P.A.: The "infodemic" of COVID-19. Arthritis Rheumatol. 72, 1806–1808 (2020). https://doi.org/10.1002/art.41468.

15. Naeem, S.B., Bhatti, R.: The Covid-19' infodemic': a new front for information professionals. Health Info. Libr. J. 37, 233–239 (2020). https://doi.org/10.1111/hir.12311.
16. Mian, A., Khan, S.: Coronavirus: the spread of misinformation. BMC Med. 18, (2020). https://doi.org/10.1186/s12916-020-01556-3.
17. Gabarron, E., Oyeyemi, S.O., Wynn, R.: COVID-19-related misinformation on social media: a systematic review. Bull. World Health Organ. 99, 455-463A (2021). https://doi.org/10.2471/blt.20.276782.
18. Thakur, N., Cui, S., Knieling, V., Khanna, K., Shao, M.: Investigation of the misinformation about COVID-19 on YouTube using topic modeling, sentiment analysis, and language analysis. Computation (Basel). 12, 28 (2024). https://doi.org/10.3390/computation12020028.
19. Vosoughi, S., Roy, D., Aral, S.: The spread of true and false news online. Science. 359, 1146–1151 (2018). https://doi.org/10.1126/science.aap9559.
20. Cinelli, M., Quattrociocchi, W., Galeazzi, A., Valensise, C.M., Brugnoli, E., Schmidt, A.L., Zola, P., Zollo, F., Scala, A.: The COVID-19 social media infodemic. Sci. Rep. 10, 1–10 (2020). https://doi.org/10.1038/s41598-020-73510-5.
21. Lazer, D.M.J., Baum, M.A., Benkler, Y., Berinsky, A.J., Greenhill, K.M., Menczer, F., Metzger, M.J., Nyhan, B., Pennycook, G., Rothschild, D., Schudson, M., Sloman, S.A., Sunstein, C.R., Thorson, E.A., Watts, D.J., Zittrain, J.L.: The science of fake news. Science. 359, 1094–1096 (2018). https://doi.org/10.1126/science.aao2998.
22. Borges do Nascimento, I.J., Beatriz Pizarro, A., Almeida, J., Azzopardi-Muscat, N., André Gonçalves, M., Björklund, M., Novillo-Ortiz, D.: Infodemics and health misinformation: a systematic review of reviews. Bull. World Health Organ. 100, 544–561 (2022). https://doi.org/10.2471/blt.21.287654.
23. Muhammed T, S., Mathew, S.K.: The disaster of misinformation: a review of research in social media. Int. J. Data Sci. Anal. 13, 271–285 (2022). https://doi.org/10.1007/s41060-022-00311-6.
24. Thakur, N.: Sentiment analysis and text analysis of the public discourse on Twitter about COVID-19 and MPox. Big Data Cogn. Comput. 7, 116 (2023). https://doi.org/10.3390/bdcc7020116.
25. Thakur, N., Cui, S., Khanna, K., Knieling, V., Duggal, Y.N., Shao, M.: Investigation of the gender-specific discourse about online learning during COVID-19 on Twitter using sentiment analysis, subjectivity analysis, and toxicity analysis. Computers. 12, 221 (2023). https://doi.org/10.3390/computers12110221.
26. Ceron, W., Gruszynski Sanseverino, G., de-Lima-Santos, M.-F., Quiles, M.G.: COVID-19 fake news diffusion across Latin America. Soc. Netw. Anal. Min. 11, (2021). https://doi.org/10.1007/s13278-021-00753-z.
27. Skafle, I., Nordahl-Hansen, A., Quintana, D.S., Wynn, R., Gabarron, E.: Misinformation about COVID-19 vaccines on social media: Rapid review. J. Med. Internet Res. 24, e37367 (2022). https://doi.org/10.2196/37367.
28. Zhao, S., Hu, S., Zhou, X., Song, S., Wang, Q., Zheng, H., Zhang, Y., Hou, Z.: The prevalence, features, influencing factors, and solutions for COVID-19 vaccine misinformation: Systematic review. JMIR Public Health Surveill. 9, e40201 (2023). https://doi.org/10.2196/40201.

29. Balakrishnan, V., Ng, W.Z., Soo, M.C., Han, G.J., Lee, C.J.: Infodemic and fake news – A comprehensive overview of its global magnitude during the COVID-19 pandemic in 2021: A scoping review. Int. J. Disaster Risk Reduct. 78, 103144 (2022). https://doi.org/10.1016/j.ijdrr.2022.103144.
30. Kisa, S., Kisa, A.: A comprehensive analysis of COVID-19 misinformation, public health impacts, and communication strategies: Scoping review. J. Med. Internet Res. 26, e56931 (2024). https://doi.org/10.2196/56931.
31. Dahlstrom, M.F.: The narrative truth about scientific misinformation. Proc. Natl. Acad. Sci. U. S. A. 118, (2021). https://doi.org/10.1073/pnas.1914085117.
32. Kotseva, B., Vianini, I., Nikolaidis, N., Faggiani, N., Potapova, K., Gasparro, C., Steiner, Y., Scornavacche, J., Jacquet, G., Dragu, V., della Rocca, L., Bucci, S., Podavini, A., Verile, M., Macmillan, C., Linge, J.P.: Trend analysis of COVID-19 mis/disinformation narratives–A 3-year study. PLoS One. 18, e0291423 (2023). https://doi.org/10.1371/journal.pone.0291423.
33. Barua, Z., Barua, S., Aktar, S., Kabir, N., Li, M.: Effects of misinformation on COVID-19 individual responses and recommendations for resilience of disastrous consequences of misinformation. Prog. Disaster Sci. 8, 100119 (2020). https://doi.org/10.1016/j.pdisas.2020.100119.
34. Bin Naeem, S., Kamel Boulos, M.N.: COVID-19 misinformation online and health literacy: A brief overview. Int. J. Environ. Res. Public Health. 18, 8091 (2021). https://doi.org/10.3390/ijerph18158091.
35. Tasnim, S., Hossain, M.M., Mazumder, H.: Impact of rumors and misinformation on COVID-19 in social media. J. Prev. Med. Public Health. 53, 171–174 (2020). https://doi.org/10.3961/jpmph.20.094.
36. Hossain, T., Logan, R.L., IV, Ugarte, A., Matsubara, Y., Young, S., Singh, S.: COVIDLies: Detecting COVID-19 misinformation on social media. In: Verspoor, K., Cohen, K.B., Conway, M., de Bruijn, B., Dredze, M., Mihalcea, R., and Wallace, B. (eds.) Proceedings of the 1st Workshop on NLP for COVID-19 (Part 2) at EMNLP 2020. Association for Computational Linguistics, Stroudsburg, PA, USA (2020).
37. Yilmaz, T., Ulusoy, Ö.: Misinformation propagation in online social networks: Game theoretic and reinforcement learning approaches. IEEE Trans. Comput. Soc. Syst. 10, 3321–3332 (2023). https://doi.org/10.1109/tcss.2022.3208793.
38. Thakur, N.: Social media mining and analysis: A brief review of recent challenges. Information (Basel). 14, 484 (2023). https://doi.org/10.3390/info14090484.
39. Wu, L., Morstatter, F., Carley, K.M., Liu, H.: Misinformation in social media: Definition, manipulation, and detection. SIGKDD Explor. 21, 80–90 (2019). https://doi.org/10.1145/3373464.3373475.
40. Shahsavari, S., Holur, P., Wang, T., Tangherlini, T.R., Roychowdhury, V.: Conspiracy in the time of corona: automatic detection of emerging COVID-19 conspiracy theories in social media and the news. J. Comput. Soc. Sci. 3, 279–317 (2020). https://doi.org/10.1007/s42001-020-00086-5.
41. van der Linden, S., Roozenbeek, J., Compton, J.: Inoculating against fake news about COVID-19. Front. Psychol. 11, (2020). https://doi.org/10.3389/fpsyg.2020.566790.

42. Narra, M., Umer, M., Sadiq, S., Eshmawi, A.A., Karamti, H., Mohamed, A., Ashraf, I.: Selective feature sets based fake news detection for COVID-19 to manage infodemic. IEEE Access. 10, 98724–98736 (2022). https://doi.org/10.1109/access.2022.3206963.
43. Wang, X., Chao, F., Yu, G., Zhang, K.: Factors influencing fake news rebuttal acceptance during the COVID-19 pandemic and the moderating effect of cognitive ability. Comput. Human Behav. 130, 107174 (2022). https://doi.org/10.1016/j.chb.2021.107174.
44. Evanega, S., Lynas, M., Adams, J., Smolenyak, K.: Coronavirus misinformation: quantifying sources and themes in the COVID-19 "infodemic," http://allianceforscience.org/wp-content/uploads/2020/10/Evanega-et-al-Coronavirus-misinformation-submitted_07_23_20-1.pdf, last accessed 2025/02/13.
45. Ferreira Caceres, M.M., Sosa, J.P., Lawrence, J.A., Sestacovschi, C., Tidd-Johnson, A., Rasool, M.H.U., Gadamidi, V.K., Ozair, S., Pandav, K., Cuevas-Lou, C., Parrish, M., Rodriguez, I., Fernandez, J.P.: The impact of misinformation on the COVID-19 pandemic. AIMS Public Health. 9, 262–277 (2022). https://doi.org/10.3934/publichealth.2022018.
46. Al-Zaman, M.S.: Prevalence and source analysis of COVID-19 misinformation in 138 countries. IFLA J. 48, 189–204 (2022). https://doi.org/10.1177/03400352211041135.
47. Elhadad, M.K., Li, K.F., Gebali, F.: Detecting Misleading Information on COVID-19. IEEE Access. 8, 165201–165215 (2020). https://doi.org/10.1109/access.2020.3022867.
48. Diepeveen, S., Pinet, M.: User perspectives on digital literacy as a response to misinformation. Dev. Policy Rev. 40, (2022). https://doi.org/10.1111/dpr.12671.
49. Czerniak, K., Pillai, R., Parmar, A., Ramnath, K., Krocker, J., Myneni, S.: A scoping review of digital health interventions for combating COVID-19 misinformation and disinformation. J. Am. Med. Inform. Assoc. 30, 752–760 (2023). https://doi.org/10.1093/jamia/ocad005.
50. Conroy, N.K., Rubin, V.L., Chen, Y.: Automatic deception detection: Methods for finding fake news. Proc. Assoc. Inf. Sci. Technol. 52, 1–4 (2015). https://doi.org/10.1002/pra2.2015.145052010082.
51. Bodaghi, A., Schmitt, K.A., Watine, P., Fung, B.C.M.: A literature review on detecting, verifying, and mitigating online misinformation. IEEE Trans. Comput. Soc. Syst. 11, 5119–5145 (2024). https://doi.org/10.1109/tcss.2023.3289031.
52. ESOC COVID-19 misinformation dataset, https://esoc.princeton.edu/publications/esoc-covid-19-misinformation-dataset, last accessed 2025/02/13.
53. Briant, E.L.: Hack attacks. In: The Routledge Companion to Freedom of Expression and Censorship. pp. 285–295. Routledge, London (2023).
54. Chavda, V.P., Sonak, S.S., Munshi, N.K., Dhamade, P.N.: Pseudoscience and fraudulent products for COVID-19 management. Environ. Sci. Pollut. Res. Int. 29, 62887–62912 (2022). https://doi.org/10.1007/s11356-022-21967-4.
55. Almomani, H., Al-Qur'an, W.: The extent of people's response to rumors and false news in light of the crisis of the Corona virus. Ann. Med. Psychol. (Paris). 178, 684–689 (2020). https://doi.org/10.1016/j.amp.2020.06.011.

56. Venturini, T.: Online conspiracy theories, digital platforms and secondary orality: Toward a sociology of online monsters. Theory Cult. Soc. 39, 61–80 (2022). https://doi.org/10.1177/02632764211070962.
57. Axt, J.R., Landau, M.J., Kay, A.C.: The psychological appeal of fake-news attributions. Psychol. Sci. 31, 848–857 (2020). https://doi.org/10.1177/0956797620922785.
58. Sathya, R., Roy, S., Paul, J., Reddy, M.S.: Fake and unproven medical remedy detector using LSTM model with tensorflow framework. In: 2022 12th International Conference on Cloud Computing, Data Science & Engineering (Confluence). pp. 514–519. IEEE (2022).
59. Douglas, K.M., Sutton, R.M., Cichocka, A.: The psychology of conspiracy theories. Curr. Dir. Psychol. Sci. 26, 538–542 (2017). https://doi.org/10.1177/0963721417718261.
60. Nowak, B.M., Miedziarek, C., Pełczyński, S., Rzymski, P.: Misinformation, fears and adherence to preventive measures during the early phase of COVID-19 pandemic: A cross-sectional study in Poland. Int. J. Environ. Res. Public Health. 18, 12266 (2021). https://doi.org/10.3390/ijerph182212266.
61. Janssen, M., van der Voort, H.: Agile and adaptive governance in crisis response: Lessons from the COVID-19 pandemic. Int. J. Inf. Manage. 55, 102180 (2020). https://doi.org/10.1016/j.ijinfomgt.2020.102180.
62. O'Connor, C., Weatherall, J.O.: The Misinformation Age: How false beliefs spread. Yale University Press, New Haven, CT (2019).
63. de Ridder, J.: What's so bad about misinformation? Inquiry (Oslo). 67, 2956–2978 (2024). https://doi.org/10.1080/0020174x.2021.2002187.
64. Søe, S.O.: A unified account of information, misinformation, and disinformation. Synthese. 198, 5929–5949 (2021). https://doi.org/10.1007/s11229-019-02444-x.
65. O'Sullivan, N.J., Nason, G., Manecksha, R.P., O'Kelly, F.: The unintentional spread of misinformation on 'TikTok'; A paediatric urological perspective. J. Pediatr. Urol. 18, 371–375 (2022). https://doi.org/10.1016/j.jpurol.2022.03.001.
66. Tran, T., Valecha, R., Rad, P., Rao, H.R.: Misinformation harms: A tale of two humanitarian crises. IEEE Trans. Prof. Commun. 63, 386–399 (2020). https://doi.org/10.1109/tpc.2020.3029685.
67. Green, M., Musi, E., Rowe, F., Charles, D., Pollock, F.D., Kypridemos, C., Morse, A., Rossini, P., Tulloch, J., Davies, A., Dearden, E., Maheswaran, H., Singleton, A., Vivancos, R., Sheard, S.: Identifying how COVID-19-related misinformation reacts to the announcement of the UK national lockdown: An interrupted time-series study. Big Data Soc. 8, (2021). https://doi.org/10.1177/20539517211013869.
68. Rocha-Jimenez, T., Olivari, C., Martínez, A., Knipper, M., Cabieses, B.: "Border closure only increased precariousness": a qualitative analysis of the effects of restrictive measures during the COVID-19 pandemic on Venezuelan's health and human rights in South America. BMC Public Health. 23, (2023). https://doi.org/10.1186/s12889-023-16726-0.
69. Lee, S.K., Sun, J., Jang, S., Connelly, S.: Misinformation of COVID-19 vaccines and vaccine hesitancy. Sci. Rep. 12, 1–11 (2022). https://doi.org/10.1038/s41598-022-17430-6.

70. Bessi, A., Coletto, M., Davidescu, G.A., Scala, A., Caldarelli, G., Quattrociocchi, W.: Science vs conspiracy: Collective narratives in the age of misinformation. PLoS One. 10, e0118093 (2015). https://doi.org/10.1371/journal.pone.0118093.
71. Kassirer, J.P.: On the take: How medicine's complicity with big business can endanger your health. Oxford University Press, London, England (2004).
72. Thakur, N.: MonkeyPox2022Tweets: A large-scale Twitter dataset on the 2022 Monkeypox outbreak, findings from analysis of Tweets, and open research questions. Infect. Dis. Rep. 14, 855–883 (2022). https://doi.org/10.3390/idr14060087.
73. Scott, P.R., Mike Jacka, J.: Auditing Social Media: A governance and risk guide. John Wiley & Sons (2011).
74. Thakur, N., Cui, S., Patel, K.A., Azizi, N., Knieling, V., Han, C., Poon, A., Shah, R.: Marburg Virus outbreak and a new conspiracy theory: Findings from a comprehensive analysis and forecasting of web behavior. Computation (Basel). 11, 234 (2023). https://doi.org/10.3390/computation11110234.
75. Bublitz, M.G., Escalas, J.E., Peracchio, L.A., Furchheim, P., Grau, S.L., Hamby, A., Kay, M.J., Mulder, M.R., Scott, A.: Transformative stories: A framework for crafting stories for social impact organizations. J. Public Policy Mark. 35, 237–248 (2016). https://doi.org/10.1509/jppm.15.133.
76. Li, Z., Meng, F., Wu, B., Kong, D., Geng, M., Qiu, X., Cao, Z., Li, T., Su, Y., Liu, S.: Reviewing the progress of infectious disease early warning systems and planning for the future. BMC Public Health. 24, (2024). https://doi.org/10.1186/s12889-024-20537-2.
77. Lopreite, M., Panzarasa, P., Puliga, M., Riccaboni, M.: Early warnings of COVID-19 outbreaks across Europe from social media. Sci. Rep. 11, 1–7 (2021). https://doi.org/10.1038/s41598-021-81333-1.
78. Oeschger, T.M., McCloskey, D.S., Buchmann, R.M., Choubal, A.M., Boza, J.M., Mehta, S., Erickson, D.: Early warning diagnostics for emerging infectious diseases in developing into late-stage pandemics. Acc. Chem. Res. 54, 3656–3666 (2021). https://doi.org/10.1021/acs.accounts.1c00383.
79. Meckawy, R., Stuckler, D., Mehta, A., Al-Ahdal, T., Doebbeling, B.N.: Effectiveness of early warning systems in the detection of infectious diseases outbreaks: a systematic review. BMC Public Health. 22, (2022). https://doi.org/10.1186/s12889-022-14625-4.
80. Rogers, L., MacCormac, A.: Finding a balance: Using a pre-post test to evaluate the effectiveness of scenario based learning using a blended approach among undergraduate nursing students. Nurse Educ. Today. 147, 106573 (2025). https://doi.org/10.1016/j.nedt.2025.106573.
81. Humphery-Jenner, M.: Legislating against misinformation: Lessons from Australia's misinformation bill. Statut. Law Rev. 45, (2024). https://doi.org/10.1093/slr/hmae023.